\documentclass[aps,prb,reprint,superscriptaddress,floatfix]{revtex4-2}
\usepackage{amssymb}
\usepackage{physics}
\usepackage{graphicx}
\usepackage{times}
\usepackage{amsmath}
\usepackage{amsthm}
\usepackage{amsfonts}
\usepackage[T1]{fontenc}
\usepackage{array}
\usepackage{multirow}
\usepackage{color}
\usepackage{esint}
\usepackage{bm}
\usepackage{color}
\usepackage{bbm}
\usepackage{hyperref}
\usepackage{babel}
\newcommand{\MIPR}{\expval{\mathcal{I}}}

\newcommand{\modelname}{$t_1$-$t_2$}

\newcommand{\sLow}{s_\text{low}}
\newcommand{\sUp}{s_\text{upper}}

\usepackage{hyperref}
\usepackage[normalem]{ulem}
\hypersetup
{	colorlinks,%
	citecolor=green,%
	linkcolor=red,%
	urlcolor=blue%
}
\allowdisplaybreaks[4]
\begin{document}

\global\long\def\id{\mathbbm{1}}
\global\long\def\ui{\mathbbm{i}}
\global\long\def\ud{\mathrm{d}}

\title{Incommensurate many-body localization in the presence of long-range hopping and single-particle mobility edge}

\author{Ke Huang}
\affiliation{Department of Physics, City University of Hong Kong, Kowloon, Hong Kong SAR}
\author{DinhDuy Vu}
\affiliation{Condensed Matter Theory Center and Joint Quantum Institute, University of Maryland, College Park, Maryland 20742, USA}
\author{Xiao Li}
\email{xiao.li@cityu.edu.hk}
\affiliation{Department of Physics, City University of Hong Kong, Kowloon, Hong Kong SAR}
\affiliation{City University of Hong Kong Shenzhen Research Institute, Shenzhen 518057, Guangdong, China}
\author{S. Das Sarma}
\affiliation{Condensed Matter Theory Center and Joint Quantum Institute, University of Maryland, College Park, Maryland 20742, USA}
\date{\today}

\begin{abstract}
We study many-body localization (MBL) in the quasiperiodic \modelname\ model, focusing on the role of next-nearest-neighbor (NNN) hopping $t_2$, which introduces a single-particle mobility edge. 
The calculated phase diagram can be divided into three distinct regimes, depending on the strength of the short-range interaction $U$. 
For weak interactions ($U\ll t_1$), this model is always nonthermal. 
For intermediate interactions ($U\sim t_1$), the thermal-MBL phase transition in this model is qualitatively the same as that of the Aubry-Andre (AA) model, which is consistent with existing experimental observations. 
For strong interactions $(U\gg t_1)$, the NNN hopping produces qualitatively new physics because it breaks down the Hilbert space fragmentation present in the AA model. 
The NNN hopping is thus irrelevant when the interaction is intermediate but relevant for strong interactions. 
\end{abstract}

\maketitle
\section{Introduction}
An interacting quantum many-body system may escape thermalization if sufficiently strong disorder is present.  
This novel dynamical phase, known as many-body localization (MBL), has been studied extensively in the past decade. 
Due to the challenges of dealing with strong interaction and strong disorder simultaneously, most theoretical work on MBL relies on numerical studies in small 1D lattice models~\cite{Nandkishore2015_Review,Altman2015_Review,Abanin2019_RMP,Gopalakrishnan2020_Review}, although some rigorous results exist in 1D systems with arbitrarily strong disorder~\cite{Imbrie2016,Imbre2016}. 
Recently there have also been some debates about the stability of MBL in the thermodynamic limit~\cite{Panda2020_EPL,Suntajs2020_PRE,Sierant2020_PRL,Abanin2021_Review,Sels2021_PRE,Leblond2021_PRB,Morningstar2022_PRB,Sels2021_arXiv}, which is outside the scope of the current work. 
Nonetheless, MBL remains the only generic mechanism to break the eigenstate thermalization hypothesis~\cite{Deutsch1991,Srednicki1994} and provides crucial insights into the foundations of quantum statistical mechanics.

One of the most intriguing open questions in this field is the fate of MBL when both extended and localized degrees of freedom are present in the corresponding single-particle spectrum~\cite{Nandkishore2014_PRB,Nandkishore2015_PRB,Huse2015_PRB,Fischer2016_PRL,Nandkishore2016_Review,Hyatt2017_PRB,Lueschen2017_PRL,Bordia2017_PRX,RubioAbadal2019_PRX,An2018_PRX,An2021_PRL}. 
{This question is particularly timely given the recent experimental interests in energy-resolved MBL~\cite{Guo2020_NatPhys}.}
In this context, incommensurate lattice models with a single-particle mobility edge (SPME) are particularly relevant for studying such questions because they provide a concrete example and are relatively easy to implement in the experiment~\cite{Lueschen2017_PRL,Bordia2017_PRX,RubioAbadal2019_PRX,An2018_PRX,An2021_PRL}. 
Remarkably, it is still unclear whether the extended states can serve as an efficient bath for the localized states in the presence of (as one would expect) interactions and lead to a faster relaxation, probably because the Hilbert space spanned by these extended states is often comparable to (or even smaller than) that spanned by localized states. 
Among the numerous models with SPME~\cite{DasSarma1986_PRL,DasSarma1988_PRL,Thouless1988_PRL,Biddle2010_PRL,Biddle2011_PRB,Ganeshan2015_PRL,Li2017_PRB,Li2020_PRB}, the \modelname\ model~\cite{Biddle2011_PRB} stands out because it is close to the system implemented in recent cold atom experiments which have been shown to have an SPME~\cite{Lueschen2018_PRL,Kohlert2019_PRL}. 
Hence, a careful analysis of the \modelname\ model in the context of MBL will provide important theoretical and experimental insights into the open question of how SPME affects thermalization (or MBL) in interacting incommensurate systems. 
{In particular, we are interested in understanding whether SPME can lead to the appearance of energy-resolved MBL}. 

In this work, we study the MBL phase diagram of the \modelname\ model and compare it to that of the Aubry-Andre (AA) model, which has been studied extensively~\cite{Iyer2013_PRB,Vu2022_PRL}. 
{The central message of our work is that SPME does not survive finite interactions.} 
Specifically, for intermediate interactions $U\sim t_1$, we find that the {thermal to MBL transition} is qualitatively similar between the AA and the \modelname\ model.
This observation qualitatively explains the recent experimental finding that no fast relaxation was observed in a system with an SPME when the interaction is of order $\order{t_1}$~\cite{Kohlert2019_PRL}. 
For the noninteracting \modelname\ model, SPME emerges because the next-nearest-neighbor (NNN) hopping breaks the duality of the AA model~\cite{Biddle2010_PRL,Biddle2011_PRB,Ganeshan2015_PRL,Li2017_PRB,Li2020_PRB}. 
However, this duality is naturally broken in interacting systems because the extended phase is only characterized by a few conserved quantities while the localized phase has many more~\cite{Serbyn2013_PRL,Imbrie2017_AnnPhys}, so a direct mapping between them is unlikely.  
On the other hand, it is proposed that the MBL transition is driven by only a few parameters, i.e., typical coupling and the typical energy deviation~\cite{Roy2019}. 
As a result, the role of the $t_2$ hopping should at best be perturbative in the intermediate interaction regime, which we verify numerically.
{We believe that such a conclusion is very general and is valid for other models with an SPME as well.} 
{Meanwhile}, for strong interactions $U\gg t_1$, we demonstrate a qualitative difference between the AA and \modelname\ models.  
In particular, while the AA model dynamics for strong short-range interactions is dominated by Hilbert space fragmentation~\cite{Vu2022_PRL}, we show that the \modelname\ model does not have this feature. 
The existence of NNN hopping $t_2$ fundamentally breaks down the fragmentation structure in the strongly-interacting Hilbert space. 
This finding is a specific feature limited to the \modelname\ model only. 
Our results provide an understanding of the role of SPME and NNN hopping in the physics of incommensurate MBL. 
Importantly, our numerical results span the weak, intermediate, and strong interaction regimes, providing the complete physics of the reentrant process where the NNN hopping in the single-particle model is rendered irrelevant at intermediate interactions but becomes qualitatively important again at strong interactions as it is for the $U=0$ noninteracting model. 

\section{Model}
We consider the following fermionic system on a chain of length $L$,
\begin{align}
	H=&\sum_j\left(t_1 c^\dag_{j+1}c_j+t_2 c^\dag_{j+2}c_j +\text{H.c.}\right)\nonumber\\
	&+V\sum_j \cos(2\pi q j + \phi)n_j+U\sum_jn_jn_{j+1},
\end{align}
where $t_1$ and $t_2$ are the strength of the nearest neighbor (NN) and the NNN hopping, respectively, {$q=(\sqrt{5}-1)/2$}, and $\phi$ is a random phase. 
In addition, $V$ and $U$ are the strength of the on-site quasiperiodic potential and the short-range interaction, respectively. 
Note that $t_2=0$ corresponds to the AA model with no SPME for $U=0$. 
The NN hopping $t_1$ is set as the energy unit throughout ($t_1=1$). 
Here we only consider the $t_2=1/6$ case in the main text but leave results for other $t_2$ values in the appendices. 
Finally, we adopt the open boundary condition (OBC) for calculations using tensor network methods for numerical accuracy and take the periodic boundary condition (PBC) for simulating half-filled lattices using the exact diagonalization (ED) method. 

\begin{figure}[t]
	\centering
	\includegraphics[width=\columnwidth]{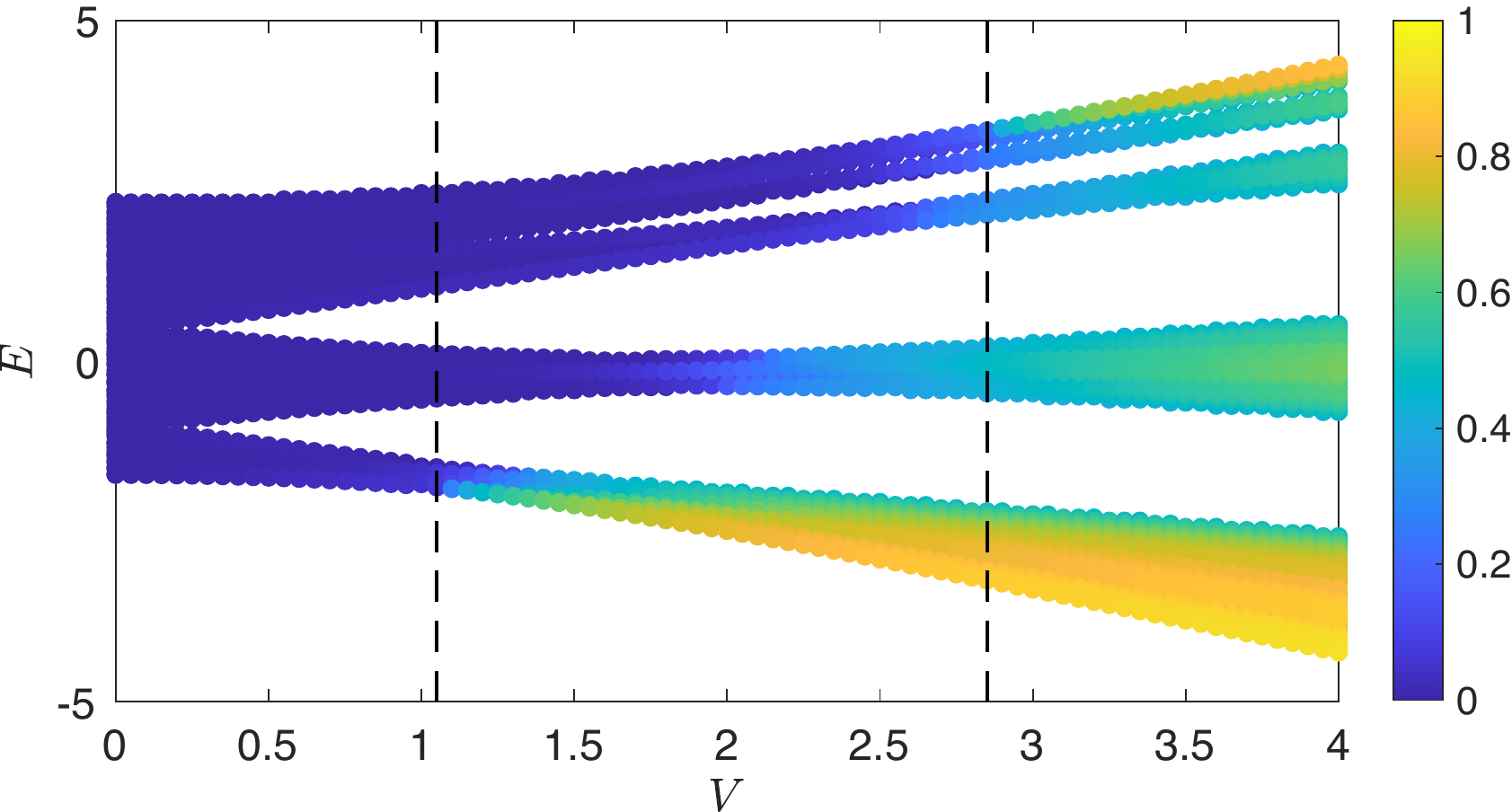}
	\caption{\label{Fig:FD} 
	{The SIPR of the eigenstates of the noninteracting \modelname\ model with $t_2=1/6$ ($t_1=1$ always) and $L=1000$.} 
	The color represents the SIPR $\expval{I_s}$, and the two dashed lines mark the range of SPME in this model. 
	In contrast, the AA model (i.e., $t_2=0$) has a single localization transition at $V=2$ without an SPME. }
\end{figure}

We first review the single-particle physics of these two models. 
The single-particle eigenstates of the AA model are localized for $V>2$~\cite{AAModel_1980}, while for the \modelname\ model, there exists an SPME~\cite{Biddle2011_PRB}. 
To quantify the extent of localization, we introduce the single-particle inverse participation ratio (SIPR) of a single-particle state $\ket{\psi}$ as $I_s =\sum_j\abs{\braket{j}{\psi}}^4$, where $\ket{j}$ is the single-particle state completely localized on site $j$. 
Hence, $I_s$ is close to $0$ for an extended state and $I_s>0$ for a localized state. 
In Fig.~\ref{Fig:FD}, we plot the SIPR of all eigenstates in the \modelname\ model with $t_2=1/6$, which indicates that localized states start to appear at $V=1.05$ and that all states are localized at $V=2.85$ (thus, an SPME for $1.05<V<2.85$; see Fig.~\ref{Fig:FD}). 
Note that though $t_2=1/6$ is relatively small, its effect on the single-particle spectrum is profound since for $t_2=0$, all states are extended up to $V=2$ in the AA model. 
In fact, we estimate that when $t_2\ll t_1$, the width of the SPME region (in terms of $V$) is about $10.4t_2$; see Appendix~\ref{Appendix:A}.

\section{Interacting phase diagram}
To characterize the MBL phase diagram of these two models, we evaluated two diagnostics: the many-body inverse participation ratio (MIPR) and the mean gap ratio. 
The MIPR of a many-body eigenstate $\ket{\psi}$ is defined as~\cite{Vu2022_PRL}
\begin{align}
	\mathcal I=\frac1{1-\nu}\qty[\frac{1}{\nu L}\sum_{i=1}^L\bar n_i^2-\nu],
\end{align}
where $\bar n_i$ is the average particle number on site $i$ and $\nu=1/2$ is the filling factor. 
Essentially, MIPR describes the particle distribution in real space. 
One can show that $\mathcal I\to 0$ for an extended state and $\mathcal I \to 1$ for a localized state. 
Hence, the mean MIPR $\MIPR$ (averaged over all eigenstates) characterizes the localization properties of a system at infinite temperature. 
Note that because our study spans both strong and weak interactions, we cannot simply examine the middle of the energy spectrum, as shown in Appendix~\ref{Appendix:B}.

\begin{figure*}[t]
\centering
	\includegraphics[width=0.7\textwidth]{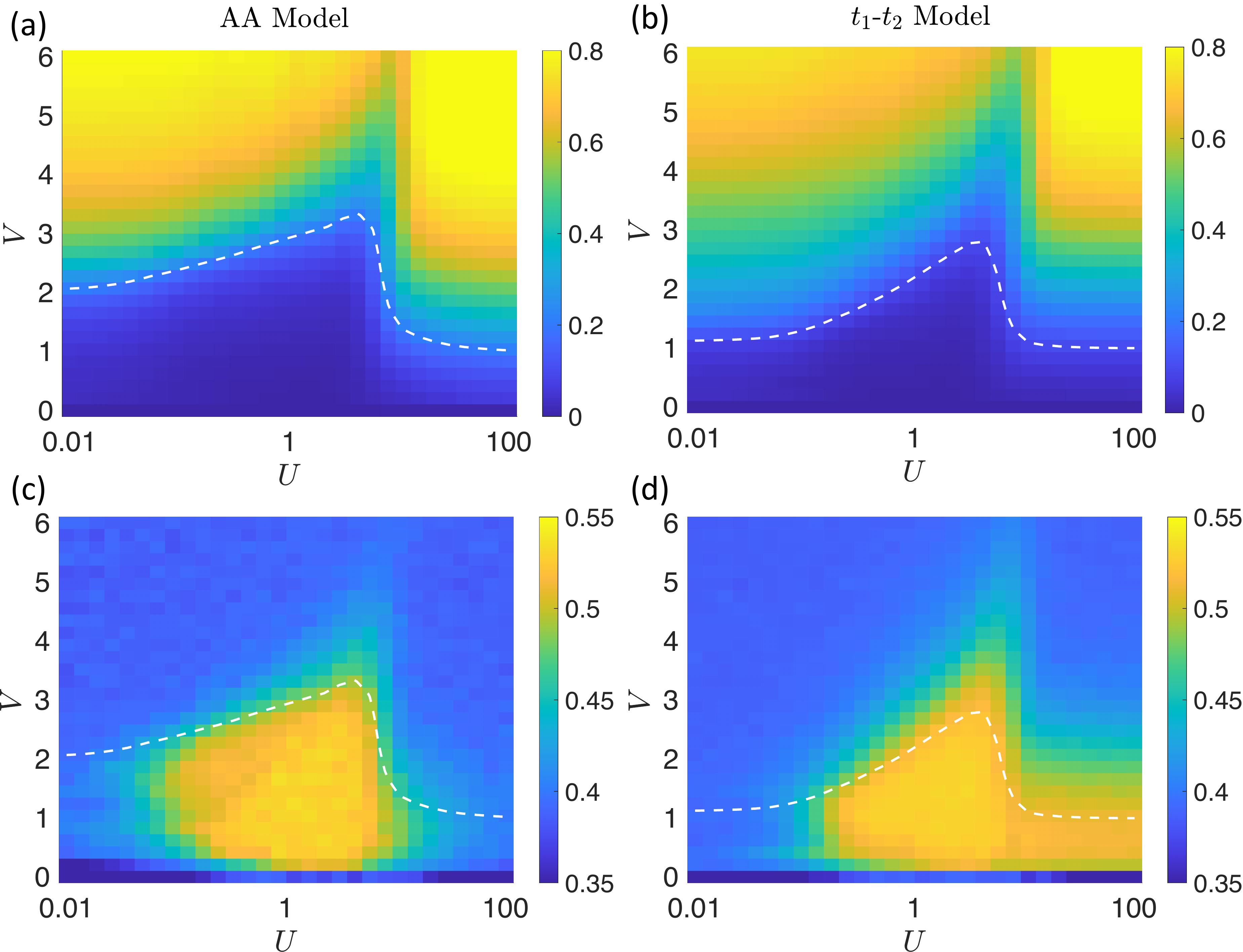}
	\caption{\label{Fig:MBL}
	Left panel: The MIPR and the mean gap ratio of the AA model. Right panel: The MIPR and the mean gap ratio of the \modelname\ model with $t_2=1/6$. The color represents the corresponding quantities. 
	Here we consider an $L=16$ system at half filling. 
	The dashed lines mark the contour for $\MIPR = 0.21$ in the AA model and $\MIPR = 0.10$ in the \modelname\ model. 
	These two values of $\MIPR$ are chosen because they correspond to the point when localized states start to appear in each model when $U=0.01$. }
\end{figure*}

In addition, we use the mean gap ratio to distinguish the thermal and MBL phases. 
The gap ratio is defined as  
$r_i=\min\qty{\frac{\delta E_i}{\delta E_{i+1}},\frac{\delta E_{i+1}}{\delta E_{i}}}$, 
where $\delta E_i=E_{i+1}-E_{i}$ is the energy gap between two adjacent eigenenergies, and the mean gap ratio $\expval{r}$ is averaged over all eigenstates. 
In the thermal phase, the spectrum follows the Gaussian orthogonal ensemble (GOE) with $\expval r=0.53$, whereas in the MBL phase, it obeys the Poisson distribution with $\expval r=0.38$. 
In Fig.~\ref{Fig:MBL}, we compare the MBL phase diagrams of the AA model and \modelname\ model. 
Both MIPR [Figs.~\ref{Fig:MBL}(a) and~\ref{Fig:MBL}(b)] and the mean gap ratio [Figs.~\ref{Fig:MBL}(c) and~\ref{Fig:MBL}(d)] indicate that the phase diagram can be clearly divided into three regimes: 
the weakly interacting ``single-particle'' regime ($U<U_{c1}$), the intermediate interaction ``many-body'' regime ($U_{c1}<U<U_{c2}$), and the strongly interacting ``Mott'' regime ($U>U_{c2}$). 
We estimate that $U_{c1}\sim0.1$ and $U_{c2}\sim10$. 
For more details on why MIPR can distinguish the three distinct regimes of these two models, including the scaling analysis of the MIPR, see Appendix~\ref{Appendix:B}. 
Additionally, in the Mott regime, where the Hilbert space is energetically partitioned, the existence of an unexpected dynamics in the \modelname\ model is a new qualitative finding.

In the single-particle regime, the interaction is too weak to thermalize the system.
Consequently, we always have $\expval r=0.38$ for all $V$ and weak interactions. 
By contrast, $\MIPR$ can still identify the localization transition as $V$ increases. 
We approximately determine the effective finite-size localization transition point by drawing a contour with $\MIPR = \mathcal{I}_0$, where $\mathcal{I}_0$ is the mean MIPR in each model where localized states start appearing in the noninteracting spectrum ($V=2$ for AA model and $V=1.05$ for the \modelname\ model). 
{Note that in the noninteracting limit ($U=0$) MIPR equals the scaled SIPR, i.e.,  
$\expval{\mathcal I}=\frac{L\expval{I_s}-1}{L-1}$~\cite{Vu2022_PRL},} in which case $\MIPR>0$ unless all single-particle states are extended. 
Therefore, we expect $\mathcal{I}_0 = 0$ in the thermodynamic limit. 
For both models, we find that in this single-particle regime, the critical $V$ does not change notably as $U$ increases. 
Hence, this regime is dominated by the single-particle properties as discussed in the previous section.

The many-body regime, by contrast, possesses a much richer structure than the weakly interacting single-particle regime. 
First, the thermal phase appears in the many-body regime as indicated by $\expval r=0.53$. 
Second, $\MIPR$ manifests a behavior similar to that of $\expval r$, with $\MIPR>0$ indicating the MBL phase.  
In particular, at $U=U_{c1}$, the thermal phase is continuously connected to the single-particle extended phase, indicating that the system is thermal only if all single-particle states are extended. 
As a result of their different single-particle localization properties (i.e., existence or not of SPME), the two models are quite different {at $U = U_{c1}$}. 
However, as $U$ increases, the localized single-particle states become thermalized, causing the thermal phase to expand. 
Meanwhile, the difference between the localization transition points of the two models decreases, {as indicated by the MIPR and mean gap ratio results in Fig.~\ref{Fig:MBL}}. 
Eventually, despite their drastically different single-particle properties, these two models behave quite similarly for $U\sim\order{1}$ in the many-body regime. 
For example, when $U=1$, both models have an MBL transition at $V\sim 3.3$, as shown in Appendix~\ref{Appendix:C}.
This intermediate $U\sim \order{1}$ regime is neither a weakly interacting perturbative nor strongly interacting Mott regime, leading to both models behaving similarly. 
{The similarity between these two models at $U\sim\order{1}$} is further verified below by the imbalance dynamics and the butterfly velocity results. 
Finally, the boundary determined previously by the contour of $\MIPR$ also qualitatively delineates the boundary of the MBL phase, although it cannot completely describe the behavior of $\MIPR$ and $\expval r$ near $U=U_{c2}$.

\begin{figure}[t]
	\includegraphics[width=\columnwidth]{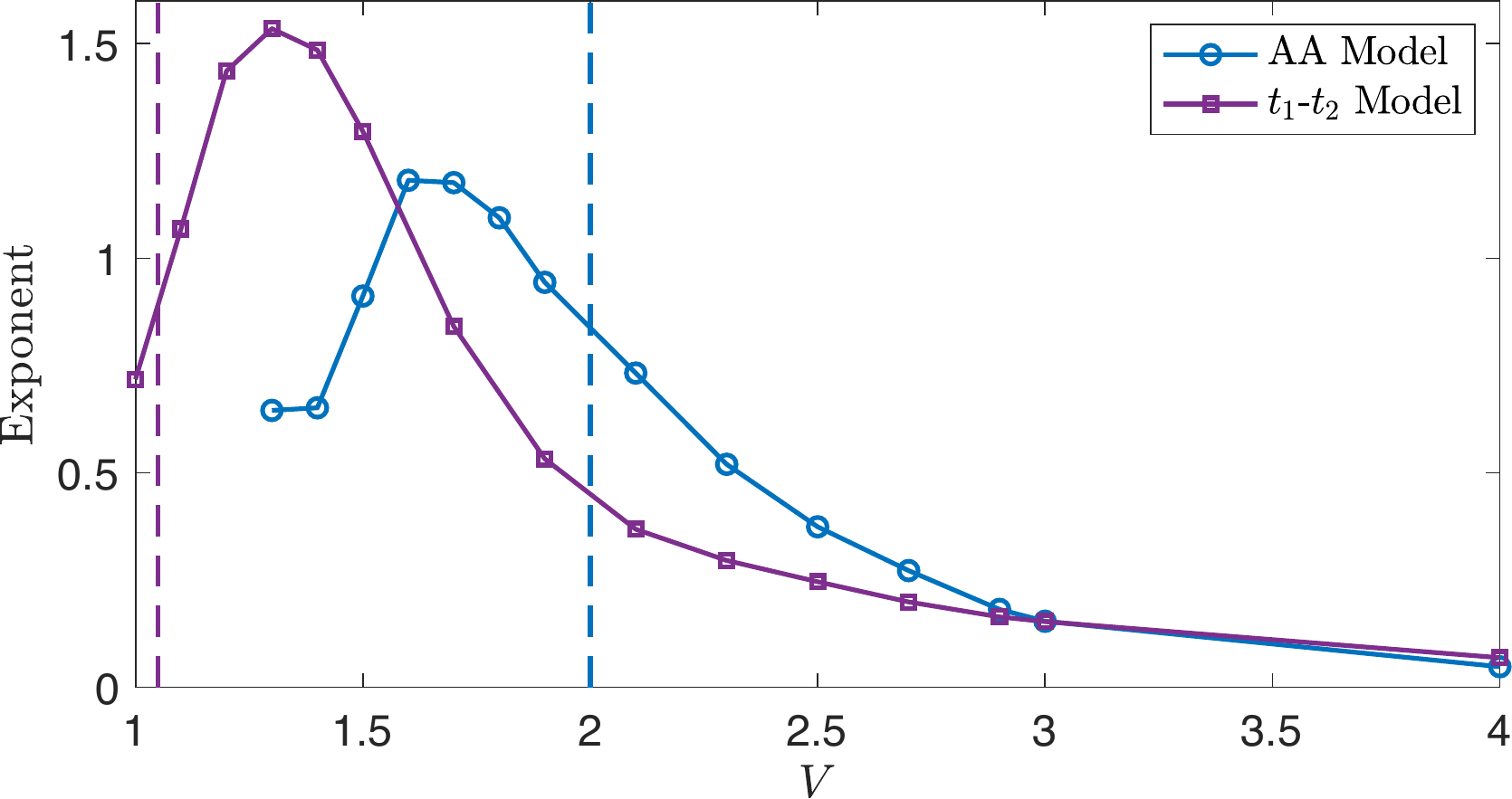}
	\center
	\caption{\label{Fig:Exp} Exponent $\alpha$ of the imbalance dynamics of the N\'{e}el state for the AA and the \modelname\ model with $t_2=1/6$.
	Here we take $U=1$, $L=24$ and the imbalance is averaged over 20 phase realizations.}
\end{figure}

\subsection{Imbalance dynamics}
We now provide additional evidence that the AA model and the \modelname\ model are qualitatively similar in the many-body regime. 
We first consider the imbalance dynamics around the thermal-MBL transition. 
A hallmark of this transition is a slow dynamics regime~\cite{Lueschen2017_PRL,Kohlert2019_PRL} in the AA model such that typical initial density profiles relax algebraically slowly to the equilibrium. 
This slow relaxation can be probed by the imbalance dynamics starting from the N\'{e}el state {$\prod_jc^\dag_{2j}\ket{0}$}, given by $I(t)=2\sum_j (-1)^j\bar{n}_j(t)/L$, 
{where $\bar{n}_j(t)$ is the expectation value of particle density at site $j$ at time $t$.} 
This process can be simulated efficiently by the KPM method~\cite{Weisse2006_RMP}. 
In the thermal phase the imbalance decays exponentially fast to zero, while in the MBL phase the imbalance remains close to one. 
By contrast, the imbalance exhibits a power-law decay in the slow-dynamics regime. 
Therefore, by fitting the imbalance to $I(t)\propto t^{-\alpha}$, we can extract the exponent $\alpha$, whose peak can be regarded as the onset of the slow-dynamics regime. 
Hence, a natural question is what happens for the \modelname\ model, which has an SPME. 
In Fig.~\ref{Fig:Exp}, we compare the exponent $\alpha$ of the AA model with that of the \modelname\ model. 
The interaction strength is $U=1$ for both models so that the system is not just trivially connected to the single-particle regime.
We find that the peak of $\alpha$ occurs at $V\approx1.6$ for the AA model and $V\approx1.3$ for the \modelname\ model.  
If we compare the latter to the $V=1.05$ point where single-particle localized states first appear in the \modelname\ model, we find that the onset of the slow dynamics regime is less sensitive to the NNN hopping. 
We further verify this point by studying systems with different $t_2$, as shown in Appendix~\ref{Appendix:C}.

\subsection{Butterfly velocity}
To further demonstrate that NNN hopping does not play an essential role in the many-body regime, we compare these two models from the quantum information scrambling perspective. 
In particular, we study the out-of-time-ordered correlator (OTOC)~\cite{Xu2019_PRR,Xu2019_NatPhys}
\begin{align}\label{Eq:OTOC}
	C_x(r,t)=\frac1{2^L}\trace\left([n_x(t),n_{x+r}][n_x(t),n_{x+r}]^\dag\right),
\end{align}
where $n_x(t)$ is the particle number operator of site $x$ in the Heisenberg picture. 
The early growth of OTOC can be fitted to the following form: 
\begin{align}
	C_x(r,t)\propto \exp\left[-a(r-v_Bt)^{1+p}/t^p\right],
	\label{Eq:butterfly}
\end{align}
where $v_B$ is the butterfly velocity {and $a,p$ are fitting parameters}. 
When $v_B>0$, there exists a ballistic spread of quantum information in the system. 
Therefore, tracking the point when $v_B$ drops to zero can provide useful diagnostics on the onset of slow dynamics. 
In a system of $L=201$ sites with OBC, we take $x=0$ without loss of generality and set $20<r<180$ in our calculation to minimize the boundary effects.

\begin{figure}[t]
	\includegraphics[width=\columnwidth]{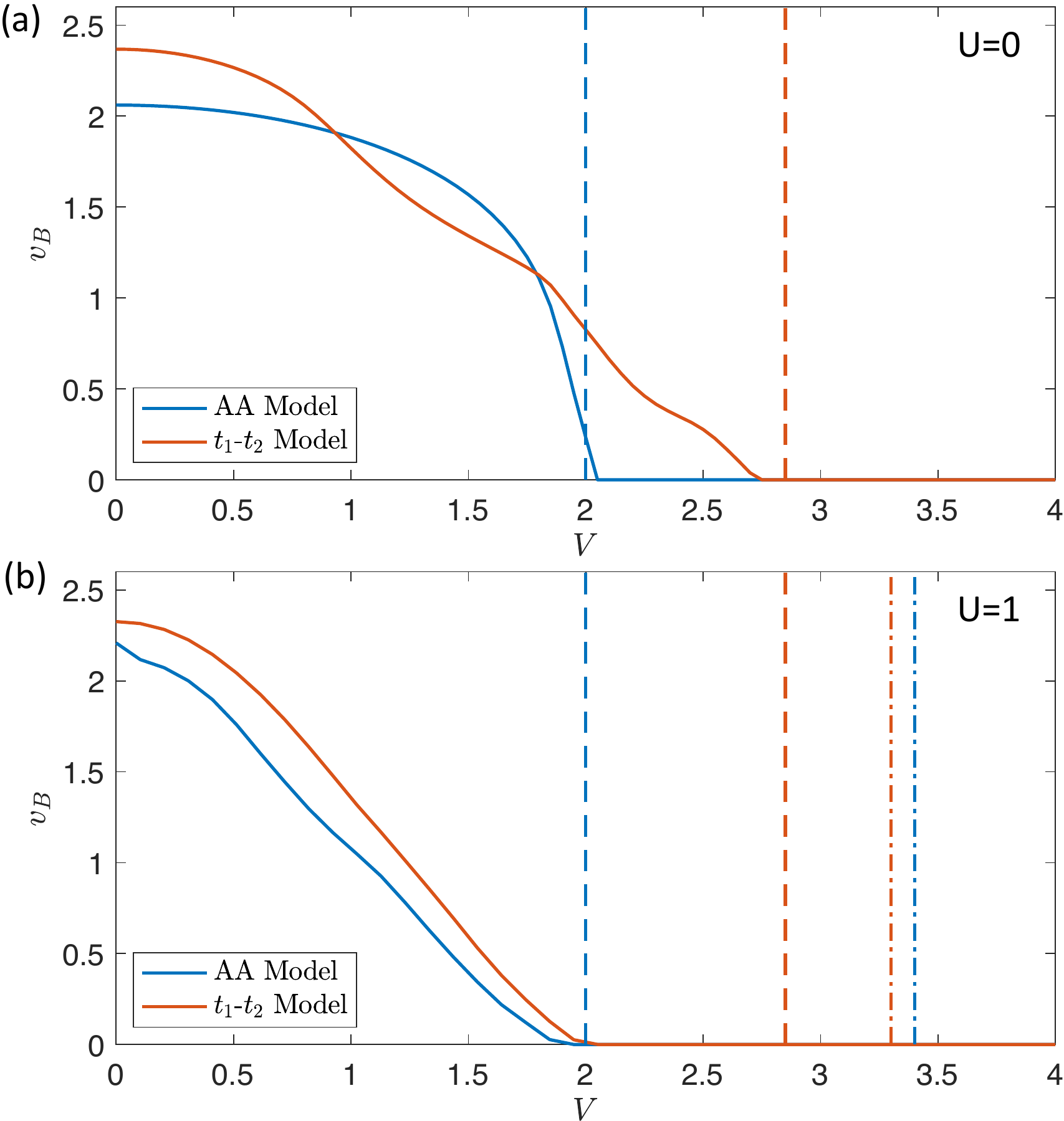}
	\center
	\caption{\label{Fig:Butterfly}
	Butterfly velocity for the AA model and the \modelname\ model with $t_2=1/6$. 
	The dashed vertical lines indicate the corresponding single-particle critical $V$ where all single-particle states are localized. 
	Finally, the two dotted lines mark the MBL transition points for each model.  
	Here we take $L=201$. }
\end{figure}

We first study the OTOC in the single-particle limit as a consistency check. 
In this case Eq.~\eqref{Eq:OTOC} simplifies to 
\begin{align}
	C_x(r,t)=\frac12 u(x,x+r,t)^2\left[1-u(x,x+r,t)^2\right],
\end{align}
where $u(x,y,t)=\abs{\expval{c_ye^{-iHt}c^\dag_x}{0}}$, with $\ket{0}$ being the vacuum state. 
We note that $v_B>0$ as long as there exist extended states because any extended state can cause $c^\dag_x\ket{0}$ to spread throughout the system and hence dominate the early growth of $u(x,y,t)$.  
Furthermore, since we only require $C_x(r,t)$ to be proportional to the right-hand side of Eq.~\eqref{Eq:butterfly}, $v_B$ is always finite unless the entire single-particle spectrum is localized. 
In Fig.~\ref{Fig:Butterfly}(a), we calculate the butterfly velocity by ED for the AA and the \modelname\ model with $L=201$ in the noninteracting limit. 
Within the fitting uncertainty, we find that $v_B$ vanishes around $V=2$ for the AA model and around $V=2.85$ for the \modelname\ model, which is exactly the single-particle localization point for the respective model. 

Next, we evaluate the OTOC in the interacting system with $U=1$. 
We use the time-dependent variational principle (TDVP) method to compute the butterfly velocity in a system with $L=201$ sites. 
We choose a bond dimension of $\chi=40$, as the calculation of {the early growth of OTOC usually does not require a large bond dimension~\cite{Xu2019_NatPhys}}. 
We have also verified this fact explicitly in Appendix~\ref{Appendix:D}.  
{In addition, the time step of our calculation is $\hbar/(200t_1)$, while the time window used for the fitting is $200\hbar/t_1$.} 
Our results for the interacting OTOC are shown in Fig.~\ref{Fig:Butterfly}(b). 
Remarkably, in contrast to the noninteracting case, $v_B$ in the interacting model almost vanishes at the same point $V\approx 2$ for both models, even though the two models have very different single-particle physics. 
Interestingly, for the AA model, the $v_B=0$ diagnostics suggests the same onset of the slow-dynamics regime as the imbalance, but for the \modelname\ model, they are different. 
This difference may be attributed to finite-size effects but also can arise because the density imbalance measures charge relaxation while the OTOC measures the information propagation. 
Nonetheless, both the imbalance dynamics and the butterfly velocity suggest that the dynamics in the many-body regime is insensitive to the NNN hopping and has no connection to the SPME.

\section{The Mott regime}
In contrast to the intermediate-coupling many-body regime, the NNN hopping plays a crucial role in the Mott regime ($U > U_{c2}$), leading to important differences between the AA model and the \modelname\ model. 
In this regime, the critical $V$ for the MBL transition decreases suddenly.  
In addition, we find that while the mean MIPR $\MIPR$ can still distinguish the thermal-to-MBL transition in the Mott regime in both models [see Figs.~\ref{Fig:MBL}(a) and~\ref{Fig:MBL}(b)], we need to be more careful about interpreting the mean gap ratio $\expval{r}$ results.  
In particular, it seems that $\expval{r}$ for the \modelname\ model exhibits a clear MBL transition when $U\gtrsim U_{c2}$ [Fig.~\ref{Fig:MBL}(c)], while that for the AA model fails to show the same transition [Fig.~\ref{Fig:MBL}(d)]. 
The reason is that in the Mott regime, the Hilbert space is highly fragmented in the strongly interacting AA model~\cite{Vu2022_PRL}, and thus $\expval r$ mixes the spectrum of different invariant subspaces, resulting in a seemingly uncorrelated spectrum. 
However, the fragmentation structure in the AA model is broken down by the NNN hopping in the \modelname\ model, and the whole many-body spectrum becomes correlated. 
We develop a rigorous domain wall argument to show that even a small $t_2$ can break down the fragmentation structure in the AA model because particles can now form doublons that connect different invariant subspaces, as shown in Appendix~\ref{Appendix:E}.

\section{Discussion and Conclusion}
We present the MBL phase diagram in the \modelname\ model and compare it with the AA model. 
{The central message of our work is that SPME may not survive finite interactions in the \modelname\ model.} 
In particular, in the many-body regime, we find that the effect of $t_2$ is negligible, {which is likely the reason why the SPME does not affect the MBL physics}. 
In fact, we have verified that both the interacting AA model and the \modelname\ model have extended states concentrated in the midspectrum and localized states at the edges of the spectrum, as shown in Appendix~\ref{Appendix:F}. 
This generic feature is probably related to the Lifschitz tail in the Anderson model and not a direct extension of the SPME. 
Therefore our work shows that the many-body mobility edge, even if it may exist in certain cases, has no connection to the SPME. 
This conclusion is likely very general and applicable to other models with an SPME. 
The other finding of our work is that, in the Mott regime, $t_2$ gives rise to qualitatively new physics: it breaks down the Hilbert space fragmentation structure in the AA model, leading to an extended thermal region. 
In fact, one can show that the thermal region in the \modelname\ model grows with increasing $t_2$ in the Mott regime, see Appendix~\ref{Appendix:E}. 
This difference can be attributed to how the many-body Hilbert space is connected by different hopping processes ($t_1$ vs $t_2$), as shown in Appendix~\ref{Appendix:F}. 
However, we emphasize that this conclusion is specific to the \modelname\ model. 
Our predictions can be verified by experiments using ultracold atoms. 
In particular, we anticipate that our results are valid even if longer-range hopping terms are present (beyond NNN), which are inevitable in these experiments.

\section*{Acknowledgements}
This work is supported by Microsoft and Laboratory for Physical Sciences. 
This work is also generously supported by the High-Performance Computing Center (HPCC) at the University of Maryland. 
K.H. is supported by the Hong Kong PhD Fellowship Scheme. 
X.L. also acknowledges support from the National Natural Science Foundation of China (Grant~No.~11904305), 
the Research Grants Council of Hong Kong (Grants~No.~CityU~21304720, No.~CityU~11300421, and No.~C7012-21G), 
as well as City University of Hong Kong (Project~No.~9610428). 

K.H. and D.V. contributed equally to this work. 

\appendix

\section{The width of SPME region for small $t_2$ \label{Appendix:A}}
In the main text, we have shown that a small $t_2$ can give rise to a wide SPME region in the \modelname\ model. 
Hence, one may think that the critical $V$ is singular at $t_2=0$.  
After all, the NNN hopping $t_2$ breaks the self-duality of the AA model under Fourier transformation, which may induce an abrupt change.  
However, as we now prove, the critical $V$ has a finite derivative at $t_2=0$.

To start with, we consider a long-range hopping model
\begin{align*}
    H=\sum_{j}\sum_{d\geq 1}t_2^{d-1}\left(c^\dag_{j+d}c_j+\text{H.c.}\right)+V\sum_j \cos(2\pi q j)n_j,
\end{align*}
which is known to possess an exact mobility edge~\cite{Biddle2010_PRL}
\begin{align}
    E=\frac{(1+t_2^2)V-2}{2t_2}. \label{Eq:ExactME}
\end{align}
Note that here we have again set the nearest-neighbor hopping term $t_1$ as the energy unit. 
The only difference between this model and the \modelname\ model is the long-range hopping on the order of $\order{t_2^2}$. 
This difference is thus irrelevant for the derivative at $t_2=0$ and we can study the long-range model instead. 
Following Eq.~\eqref{Eq:ExactME}, we know the onset of the localization $V$ is determined by
\begin{align}
    2t_2f(V,t_2)=(1+t_2^2)V-2,
\end{align}
where $f(V,t_2)$ is the lower bound of the energy spectrum as a function of $V$ and $t_2$. 
After collecting terms on the order of $\order{t_2^2}$, we obtain
\begin{align}
    V=2+2t_2f(V,0)+\order{t_2^2},
\end{align}
and therefore we have 
\begin{align}
    \eval{\dv{V}{t_2}}_{t_2=0}=2f(V,0).
\end{align}
The above result indicates that the derivative is twice the lower bound of the spectrum of the AA model at $V=2$. 
Numerically we find that $f(V=2,t_2=0) = -2.6$ in the AA model, and thus the width of the SPME is approximately 
\begin{align}
\delta V=10.4t_2. \label{EqSM:SPME_Width}
\end{align} 
We have verified that this approximation agrees with the numerical result for $t_2=1/6$ very well.

\begin{figure}[!]
	\includegraphics[width=\columnwidth]{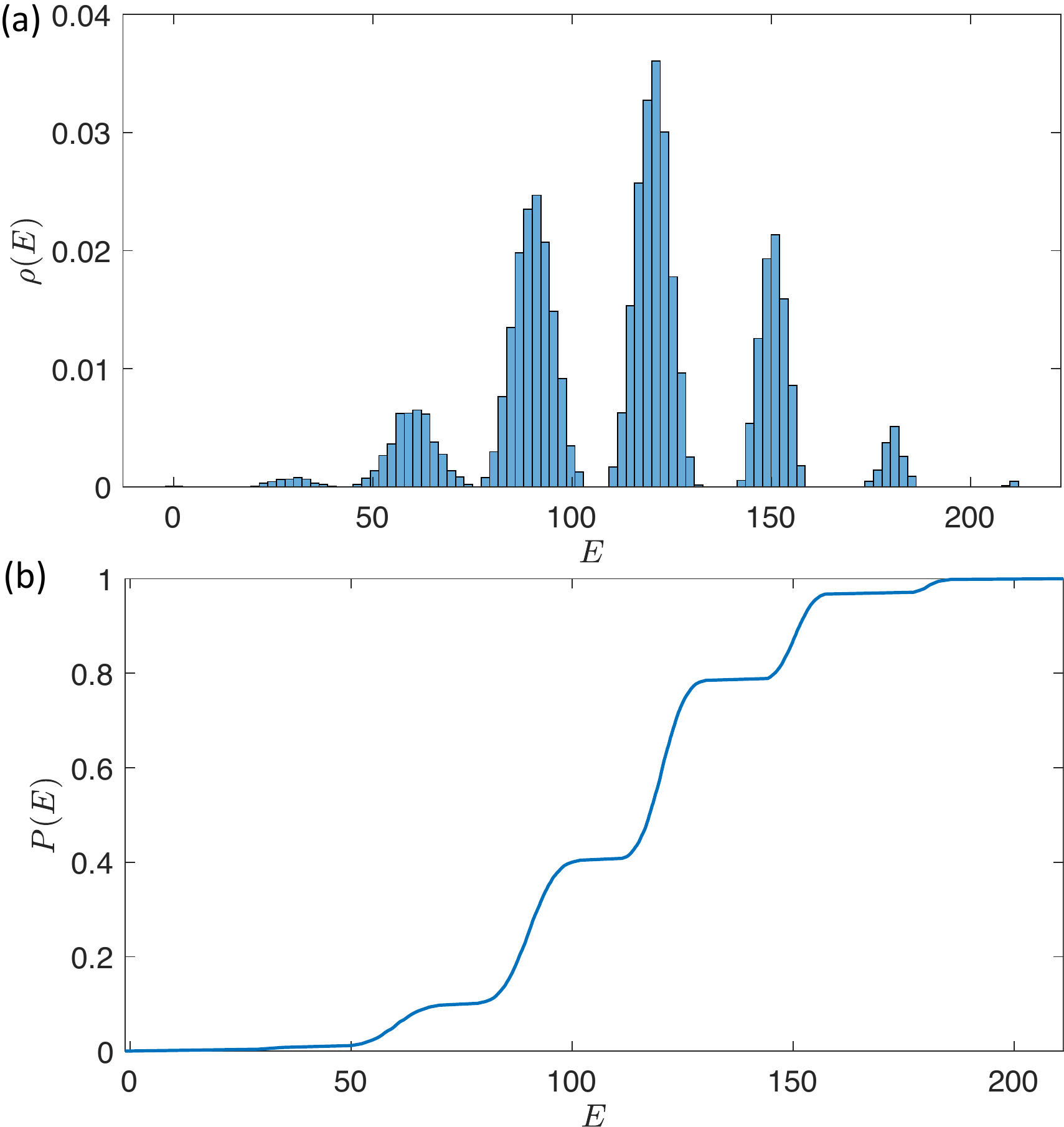}
	\center
	\caption{\label{Fig:DOS} (a) Density of states (DOS) of the interacting AA model with $L=16$, $V=2.5$, and $U=30$. (b) The integrated DOS of the same model as (a).}
\end{figure}

\section{Scaling of the MIPR \label{Appendix:B}}
\begin{figure}[!]
	\includegraphics[width=\columnwidth]{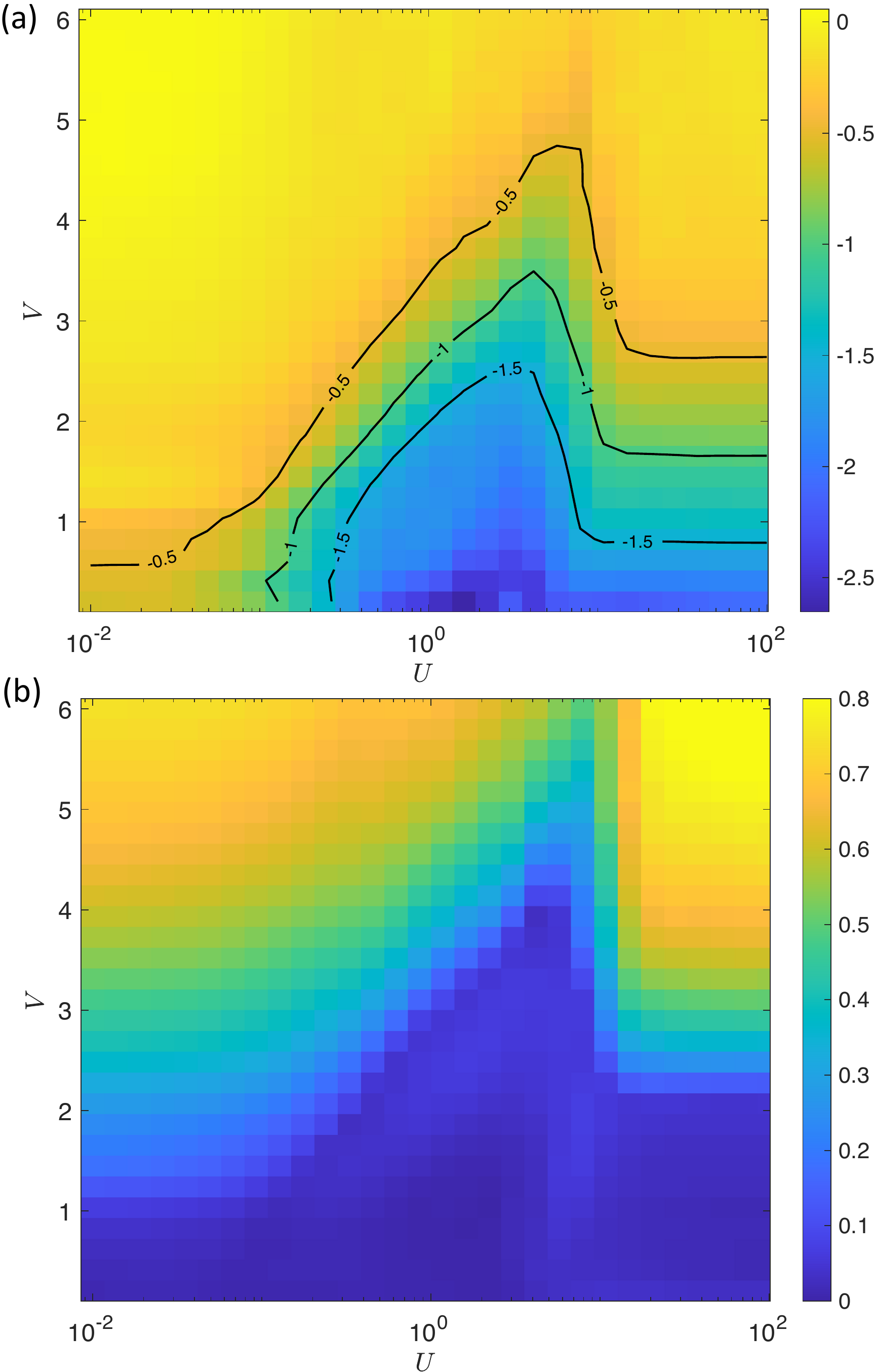}
	\center
	\caption{(a) Scaling exponent $\kappa$ and (b) extrapolated IPR $\mathcal{I}_\infty$ of the \modelname\ model with $t_2=1/6$.\label{Fig:IPR}} 
\end{figure}

\begin{figure*}[!]
	\includegraphics[width=\textwidth]{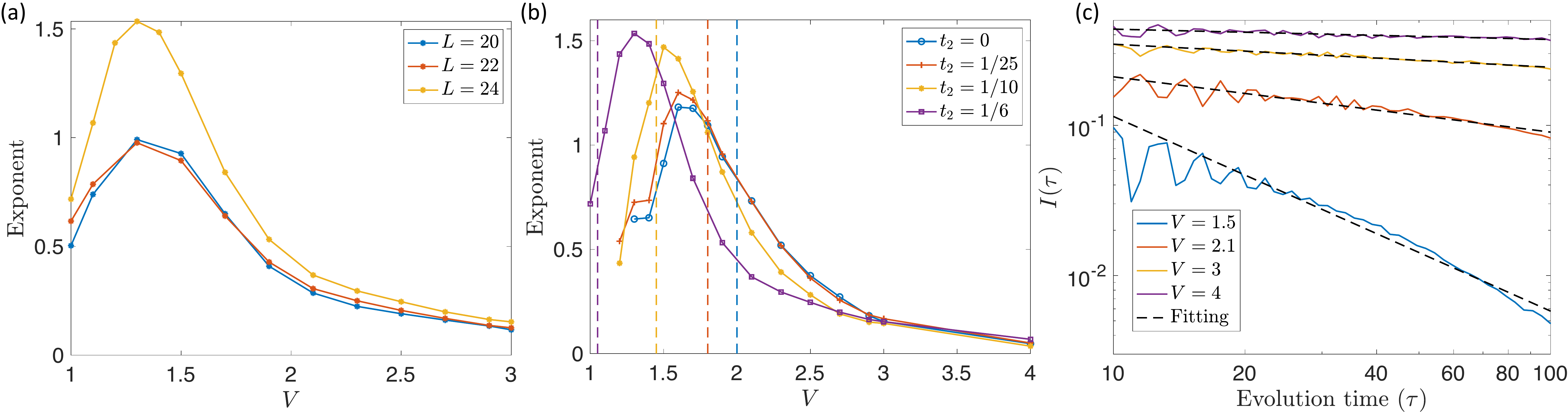}
	\center
	\caption{\label{FigSM:Exp}(a) Exponents for $t_2=1/6$ with various system sizes. (b) Solid lines are the exponents extracted from the imbalance dynamics in a $L=24$ system, and the dashed vertical lines are the critical points $V_c$ where single-particle localized states start to appear in the corresponding noninteracting limit. (c) Solid lines are the imbalance dynamics for $t_2=1/6$ and $L=24$, and the dashed lines are the fitting. Here we take $U=1$, and average over 20 phase realizations. }
\end{figure*}

In this section, we study the finite-size scaling of the MIPR. 
To begin with, we emphasize the importance of studying the entire energy spectrum instead of just focusing on the middle of the spectrum. 
The reason is that, in the presence of strong interactions, the many-body spectrum in the interacting AA model in any finite system will split into several bands according to the number of domain walls, as shown explicitly in Fig.~\ref{Fig:DOS}(a). 
Compared with Fig.~\ref{figMBME} below (where $U\sim1$), the spectrum in the Mott regime is clearly modified. 
We further calculate the integrated DOS $P(E)=\int_{-\infty}^E \rho(\epsilon)\dd \epsilon$ in Fig.~\ref{Fig:DOS}(b), which shows that all the bands, including the one in the center, share only a finite fraction of the states. 
Hence, the spectrum is by no means dominated by states in the middle of the spectrum, and focusing only on the middle in the Mott regime is misleading. 

Following the analysis in Ref.~\cite{Vu2022_PRL}, we adopt the fitting equation
$\expval{\mathcal{I}}/(1-\expval{\mathcal{I}})\propto N^\kappa$,
where $N$ is the particle number and the fitting parameter $\kappa$ is the scaling exponent. In the localized regime, $\expval{\mathcal{I}}$ does not scale with the system size, and therefore $\kappa\approx0$. 
Additionally, both the thermal and single-particle extended regimes have a nonvanishing scaling exponent. 
In contrast, the thermal regime generally has a smaller MIPR and more negative scaling exponent than the single-particle extended regime, as shown in the main text and Fig.~\ref{Fig:IPR}(a).

\begin{figure}[b]
	\includegraphics[width=\columnwidth]{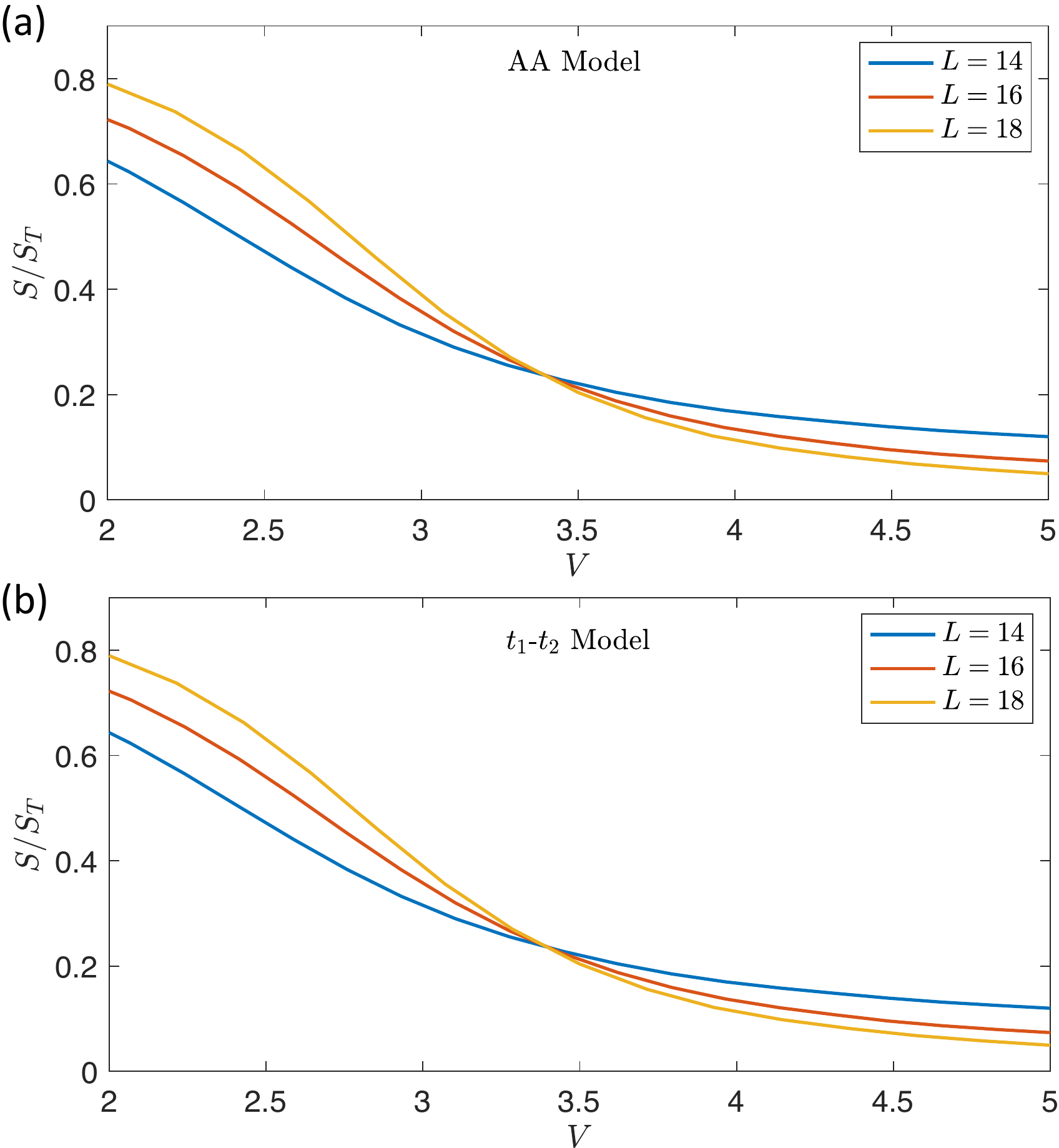}
	\center
	\caption{Entanglement entropy of the AA model and the \modelname\ model with $t_2=1/6$. 
	Additionally, we set $U=1$ for both models. 
	The results are averaged over $1000$, $200$, and $5$ random phase realizations for a system size of $L=14$, $16$, and $18$, respectively.}\label{Fig:EE}
\end{figure}

Further, utilizing the scaling exponent, one can extrapolate the MIPR in the thermodynamic limit according to the following piecewise fitting~\cite{Vu2022_PRL},
\begin{align}
	\frac{\expval{\mathcal{I}}}{1-\expval{\mathcal{I}}}= 
\begin{cases}
	a_1/N+\mathcal{I}_\infty, & 0>\kappa>-1\\
	a_2 N^\kappa+a_1/N+\mathcal{I}_\infty, & \kappa<-1
\end{cases}, 
\end{align}
where $a_1,a_2$ are the fitting parameters, and $\mathcal{I}_\infty$ is the extrapolated IPR. 
We present the result in Fig.~\ref{Fig:IPR}(b).

\section{Effects of $t_2$ on the MBL phase diagram at intermediate interactions \label{Appendix:C}}
In this section, we provide additional data on the AA model and the \modelname\ model at intermediate $U$. 
In particular, we set $U=1$ for both models, and study the entanglement entropy (EE) and the density imbalance.

\subsection{Density imbalance}
{
Recall from the main text that the imbalance results are fitted to the functional form $I(t)\propto t^{-\alpha}$. 
We first show in Fig.~\ref{FigSM:Exp}(a) that there is no simple monotonic scaling for the relaxation exponent $\alpha$ as the system size increases from $L=20$ to $L=24$. 
Notwithstanding, the exponents of different system sizes peak around the same $V$, and manifest a similar trend with respect to $V$. 
Second, we present the exponent extracted from the imbalance dynamics for various additional $t_2$ values in Fig.~\ref{FigSM:Exp}(b), {and particularly, we provide the dynamics and the fitting for some parameters explicitly in Fig.~\ref{FigSM:Exp}(c) to show the typical behavior of the imbalance.}
Remember from the main text that the peak of the exponent can be regarded as the start of the slow-dynamics regime. 
Figure~\ref{FigSM:Exp}(b) thus demonstrates that the onset $V_s$ for the slow-dynamics regime decreases as $t_2$ increases. 
However, the decrease of $V_s$ is less significant than the decrease of the critical $V_c$ at which point localized single-particle states start to appear in the system. 
As a result, although the single-particle critical $V_c$ is greater than the slow-dynamics critical $V_s$ initially (in the AA model), it will become smaller than $V_s$ as $t_2$ increases. 

\subsection{Entanglement entropy}

To determine the MBL transition at intermediate $U$ more precisely, we study the finite-size scaling of the EE. 
Note that if a system is divided into two subsystems $A$ and $B$, the second Renyi entropy of a state $\ket{\psi}$ is defined as $S=-\ln\tr_A{\rho_A^2}$, where $\rho_A=\tr_B{\dyad{\psi}}$. 
We calculate the average half-chain EE of all eigenstates for different sizes in Fig.~\ref{Fig:EE}. 
In the thermal phase, the EE approaches the Page value $S_T=(L\ln2-1)/2$~\cite{Khemani2017_PRL}, whereas the EE does not vary with the system size in the MBL phase. 
In Fig.~\ref{Fig:EE}, curves of different $L$ cross at $V=3.4$ for the AA model and at $V=3.3$ for the \modelname\ model, suggesting that the effect of $t_2$ is rather insignificant.

\section{More details on the butterfly velocity \label{Appendix:D}}
We now provide more details regarding the evaluation of the Butterfly velocity $v_B$. 
We first discuss the OTOC calculation. 
Note that we cannot use ED to evaluate OTOC for the interacting case since it can only be carried out in a small system ($L\sim 18$), which does not provide enough data for the fitting. 
Instead, tensor network methods can evaluate the buttery velocity rather accurately. 
Because of the Lieb-Robinson bounds~\cite{Lieb1972}, the operator entanglement far outside the light cone (the early growth) is also exponentially small. 
As a result, the information in such regions can be faithfully stored with a small bond dimension~\cite{Xu2019_NatPhys}. 
Specifically, we utilize the time-dependent variational principle (TDVP) to compute the butterfly velocity in a system with $L=201$ sites. 

As mentioned in the main text, in order to extract the butterfly velocity, we fit the early growth of OTOC to the following form, 
\begin{align}
	C_x(r,t)\propto \exp\left[-a(r-v_Bt)^{1+p}/t^p\right],
\end{align}
which is Eq.~(4) in the main text. 
A subtlety here is how ``early growth'' should be defined. 
We choose to fit the regime where $\log_{10} C_x(r,t)\in[\sLow, \sUp]$ , where $\sLow$ and $\sUp$ are the lower and the upper bound of the fitting, respectively. 
Specifically, we fix $\sLow$ to be slightly larger than the machine precision, and in Fig.~\ref{Fig:Butterfly} in the main text, we take $\sUp=-14$ for the noninteracting case and $\sUp = -10$ for the interacting one. 
In practice, we find that $v_B$ only weakly depends on $\sUp$. 
For example, we present the butterfly velocity extracted using various $\sUp$ in Fig.~\ref{Fig:ButterflyAppendix}, which indeed shows that $v_B$ hardly depends on $\sUp$.

\begin{figure*}[!]
	\includegraphics[width=0.9\textwidth]{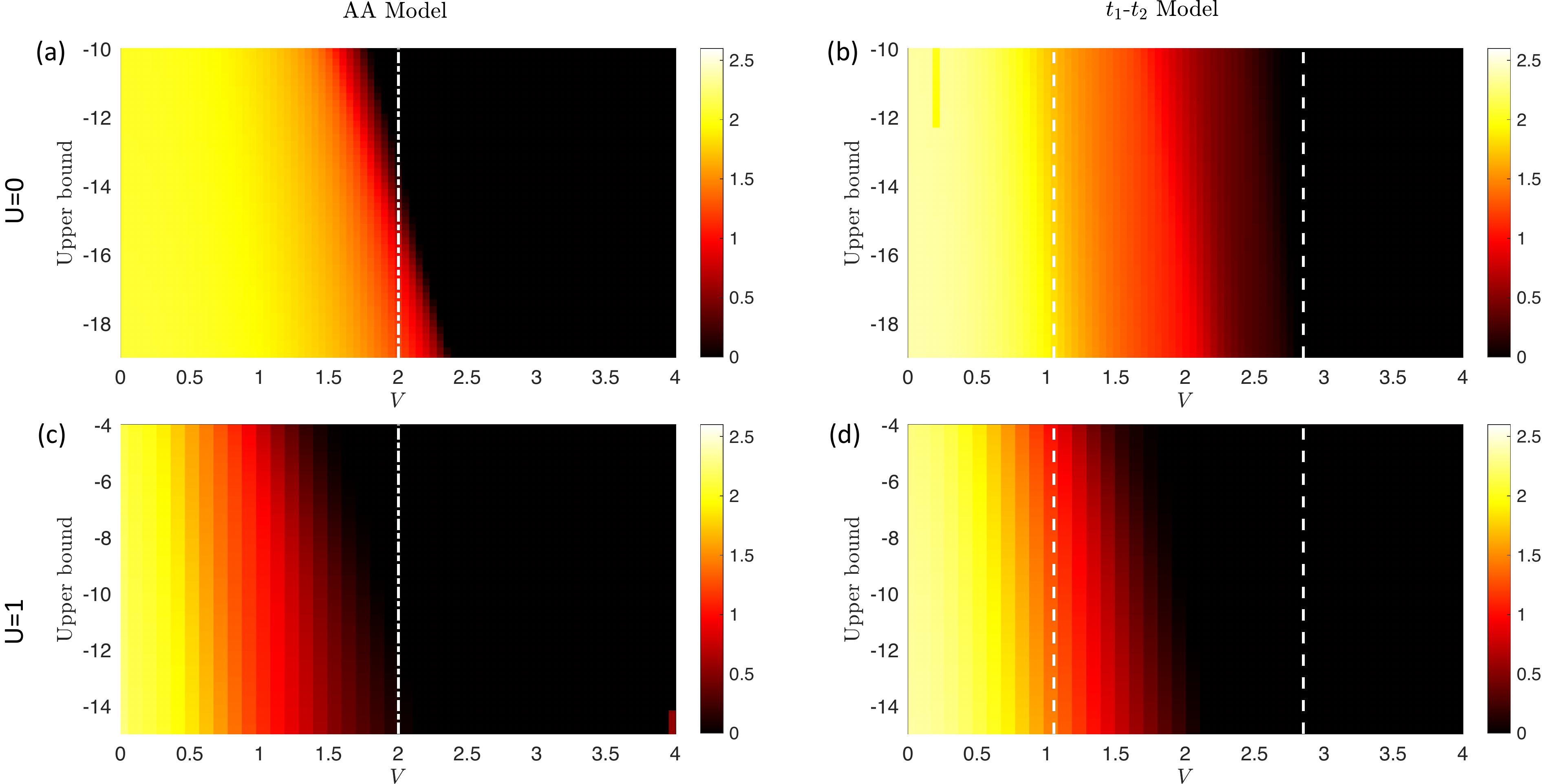}
	\center
	\caption{Butterfly velocity $v_B$ in (a) the noninteracting AA model, (b) the noninteracting \modelname\ model, (c) the interacting AA model, and (d) the interacting the \modelname\ model. 
	The color represents the butterfly velocity itself. 
	We set $t_2=1/6$ in (b) and (d), and $U=1$ in (c) and (d). The white dash-dotted lines in (a) and (c) mark the $V=2$ point (which is the single-particle localization transition in the AA model), and the white dashed lines in (b) and (d) mark the region of SPME in the \modelname\ model.}\label{Fig:ButterflyAppendix}
\end{figure*}

{
\begin{figure}[t]
\includegraphics[width=\columnwidth]{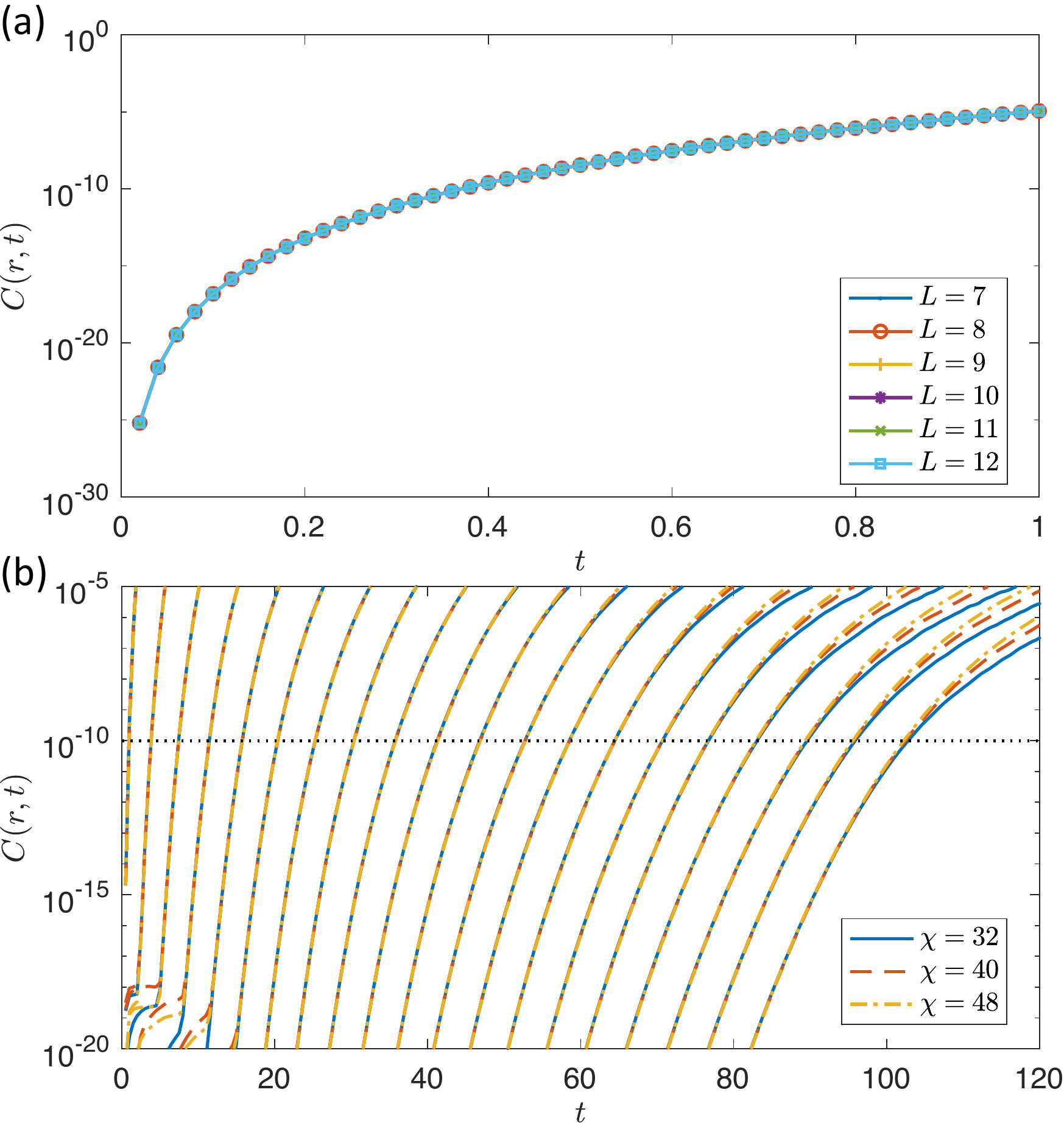}
	\center
\caption{\label{Fig:OTOC}
(a) {Convergence of OTOC in a small system}.The OTOC between site 1 and site 7 in the AA model with $U=1$ and $V=1$ for various system sizes. Note that $L=7$ is the smallest system size for the OTOC between site 1 and site 7, and even this one has almost no finite-size effect. 
(b) {Convergence of our calculation}. The OTOC in the AA model of length $L=201$ with $U=1$, $V=1$ for different bond dimensions. 
The $20$ groups of curves correspond to (from left to right) $r=10, 20,\cdots,200$.
Within each group, the three different line styles represent different bond dimensions $\chi$.}
\end{figure}

One of the most important features of the early growth of the OTOC is that it hardly suffers from finite-size effects, which we verify explicitly here. 
As an example, we use exact diagonalization to calculate the OTOC between site~1 and site~7 [i.e., evaluating $C(r=6,t)$] using various system sizes, and the result is shown in Fig.~\ref{Fig:OTOC}(a). 
It is clear from the figure that all OTOC traces agree with each other. 
The underlying reason is that the Lieb-Robinson bounds~\cite{Lieb1972} guarantee an exponentially small correction outside the light cone. 
In the early growth regime of the OTOC, the physical boundary of the lattice is far outside the light cone, and hence one cannot observe the effect of the boundary. 

Second, Second, other than the theoretical analysis on the bond dimension in~\cite{Xu2019_PRR}, we also numerically check the convergence of the calculation with respect to the bond dimension, which is shown in Fig.~\ref{Fig:OTOC}(b). 
Recall from the previous section that we extracted the butterfly velocity for the interacting AA model and \modelname\ model from the regime of $\log_{10}C_x(r,t)\in \qty[\sLow, \sUp]$, where $\sLow$ is slightly larger than the machine precision, while $\sUp=-10$. 
One can see from Fig.~\ref{Fig:OTOC}(b) that a bond dimension of $\chi=32$ is already enough to accurately simulate the early growth of OTOC in this regime, as there is no difference between $\chi=32$ and $\chi=48$. 
Therefore, our calculation with a bond dimension of $\chi=40$ should suffice. 
}

\section{Hilbert space fragmentation in the Mott regime \label{Appendix:E}}
{Hilbert space fragmentation for strongly correlated 1D systems has attracted much interest recently~\cite{Pablo2020_PRX,Vedika2020_PRB,Herviou2021_PRB}.}
For example, if there exists a strong nearest-neighbor (NN) interaction, the number of domain walls between an occupied and an empty site becomes a conserved quantity. 
Moreover, the invariant subspace is further fragmented if only NN hopping exists in the model (which is the case for the AA model). 
The same mechanism has been established in a similar model~\cite{DeTomas2019_PRB,Yang2020_PRL}, and we recapitulate it here for our discussion of the \modelname\ model.

\subsection{Hilbert space fragmentation in the AA model}
{In the AA model, the effective model is derived from the first-order pertubation theory,
\begin{align}
    H_{\text{eff}}=&t_1\sum_{j}(c^\dag_jc_{j+1}+\text{H.c.})P_{j-1,j+2}\nonumber\\
    &+V\sum_j\cos(2\pi qj+\phi)n_j.
\end{align}
where $P_{i,j}=2n_{i}n_{j}-n_{i}-n_{j}+1$ is the constraint imposed by the strong NN interaction.}
Hence, the system only allows two types of constrained hopping processes to conserve the number of domain walls, 
\begin{align}
\begin{split}
    &\bullet\circ\bullet\bullet\Longleftrightarrow\bullet\bullet\circ\bullet~\\
    &\circ\bullet\circ\circ\Longleftrightarrow\circ\circ\bullet\circ~, \label{EqSM:ConstrainedHopping}
\end{split}
\end{align}
where $\circ$ or $\bullet$ represents an empty or an occupied site, respectively. 
The first process can be regarded as a hole hopping and the second is a particle hopping. 
Hence, we notice that a single hole (or particle) always moves freely in a particle (or hole) sea, and we call it a mobile hole (or particle). 
However, when multiple holes appear in one particle sea, different holes cannot meet, i.e., 
\begin{align}
    &\bullet\bullet\circ\bullet\circ\bullet\bullet ~\nRightarrow~ \bullet\bullet\bullet\circ\circ\bullet\bullet~.
\end{align} 

Then, the question is how to distinguish the mobile part from the localized part (the sea). 
Here, we summarize the procedure to extract the sea configuration of a 1D many-body state on a ring (due to periodic boundary conditions, or PBCs) as follows: 
\begin{enumerate}
\item Find out all particles whose neighbors are both holes and all holes whose neighbors are both particles, i.e., configurations like $\circ\bullet\circ$ and $\bullet\circ\bullet$. 
Then remove these sites from the ring, which is thus divided into several disconnected fragments. 
\item Glue the edges of the fragments together. If two edges are different, we directly glue them together. 
However, if both edges are identical (i.e., both occupied or both empty), we need to remove an extra site (either of the two edges). 
\end{enumerate}

We call the resulting pattern the sea configuration of a many-body state. 
It can be readily proved that the sea configuration is invariant under the two constrained hopping processes in Eq.~\eqref{EqSM:ConstrainedHopping}.  
Note that the sea configuration should be considered in a translation-invariant way for PBCs. 
The remaining sites are mobile particle-hole pairs. 
If $N_s$ is the length of the sea configuration and $N_m$ is the number of mobile particle-hole pairs, then we have
\begin{align}
    N_s+2N_m=L.
\end{align}
Therefore, the system can be viewed as $N_m$ particles moving on a ring of length $N_s+N_m$. 
{We emphasize that the sea configuration is a structure beyond the conservation of domain walls and therefore causes the Hilbert space fragmentation~\cite{Pablo2020_PRX,Vedika2020_PRB,Herviou2021_PRB}, which is similar to the one proposed in Refs.~\cite{DeTomas2019_PRB,Yang2020_PRL}.}

\subsection{The breakdown of fragmentation by the NNN hopping $t_2$}
We now demonstrate that the NNN hopping $t_2$ can break down the fragmentation in the AA model in the Mott regime. 
Consider the following example, 
\begin{align}
    &\bullet\bullet\circ\circ\circ\circ\bullet\circ\bullet\bullet\bullet~\label{Eq:Melt1}\\
    \Longrightarrow&
    \bullet\bullet\circ\circ\circ\bullet\circ\circ\bullet\bullet\bullet~\\
    \Longrightarrow&
    \bullet\bullet\circ\circ\circ\bullet\bullet\circ\circ\bullet\bullet~\label{Eq:Melt2}\\
    \Longrightarrow&
    \bullet\bullet\circ\bullet\bullet\circ\circ\circ\circ\bullet\bullet~\label{Eq:Melt3}\\
    \Longrightarrow&
    \bullet\bullet\bullet\circ\bullet\circ\circ\circ\circ\bullet\bullet~\label{Eq:Melt4},
\end{align}
which has been presented in Fig.~5(b) in the main text. 
Here we provide additional details. 
We compare the two many-body states in Eq.~\eqref{Eq:Melt1} and Eq.~\eqref{Eq:Melt4}. 
According to our rules, the former has a sea configuration of $\bullet\bullet\circ\circ\circ\circ\bullet\bullet\bullet$, while the latter has a different sea configuration of $\bullet\bullet\bullet\circ\circ\circ\circ\bullet\bullet$. 
Without NNN hopping $t_2$ (i.e., in the AA model), these two sea configurations are dynamically disconnected due to the Hilbert space fragmentation mechanism. 

{With the inclusion of the NNN hopping $t_2$, the effective model has an additional constrained hopping,
\begin{align}\label{Eq:NNNhopping}
    \tilde H=t_2\sum_{j}(c^\dag_{j-1}c_{j+1}+\text{H.c.})P_{j-2,j+2},
\end{align}
which is able to connect different sea configurations.}
Particularly, we can show that NNN hopping can transform the state in Eq.~\eqref{Eq:Melt1} into the one in Eq.~\eqref{Eq:Melt4}. 
To see this, first note that in addition to the processes in Eq.~\eqref{EqSM:ConstrainedHopping}, Eq.~\eqref{Eq:NNNhopping} allows the \modelname\ model to move a doublon of particles in the hole sea and vice versa: 
\begin{align}
	\circ\bullet\bullet\circ\circ\circ
	\Longleftrightarrow
	\circ\circ\bullet\bullet\circ\circ.  
\end{align}
Hence, with the help of the NNN hopping, we can form a particle doublon [see Eqs.~\eqref{Eq:Melt1}-\eqref{Eq:Melt2}] at the boundary between a particle sea and a hole sea, which then moves freely in the hole sea [Eq.~\eqref{Eq:Melt3}]. 
If we compare the sea configurations of the two states in Eq.~\eqref{Eq:Melt1} and Eq.~\eqref{Eq:Melt4} (i.e., $\bullet\bullet\circ\circ\circ\circ\bullet\bullet\bullet$ vs. $\bullet\bullet\bullet\circ\circ\circ\circ\bullet\bullet$), we realize that in this example the NNN hopping effectively transfers a particle from the right particle sea to the left one.

\begin{figure}[t]
	\includegraphics[width=\columnwidth]{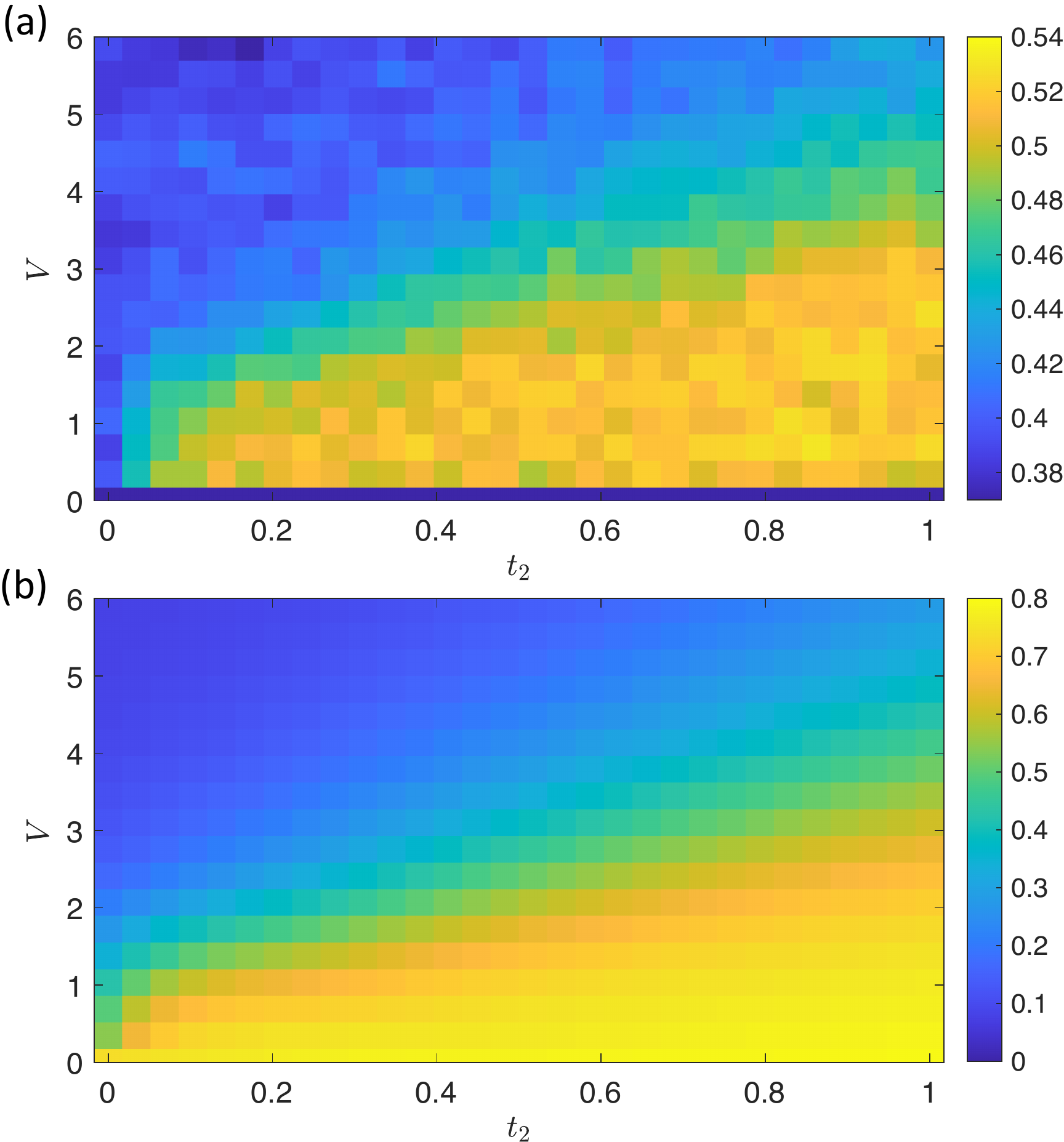}
	\center
	\caption{\label{Fig:EE&LS}
	(a) Mean gap ratio as a function of $t_2$. 
	(b) Entanglement entropy as a function of $t_2$. The color represents the corresponding quantities. We choose $U=500$, deep in the Mott regime. The system size is $L=16$ and at half filling. }
\end{figure}

The fragmentation and its breakdown result in two important phenomena in the Mott regime. 
First, the whole spectrum of the AA model resembles the Poisson distribution, while that of the \modelname\ model follows GOE when $V$ is small. 
Second, the EE in the AA model is generally lower than that in the \modelname\ model, because the dimension of the subsystem of each fragment in the AA model is less than $2^{L/2}$ (due to fragmentation)~\cite{DeTomas2019_PRB,Yang2020_PRL}. 
Both features can be observed in Fig.~\ref{Fig:EE&LS}.

Finally, we find that the thermal region in the \modelname\ model increases as $t_2$ increases, which is evident from both the level statistics and the EE result in Fig.~\ref{Fig:EE&LS}. 
Interestingly, this trend is opposite to the $U=0$ limit, where the region of the purely extended phase actually shrinks as $t_2$ increases, as shown in Eq.~\eqref{EqSM:SPME_Width}. 

\begin{figure*}[t]
	\includegraphics[width=0.8\textwidth]{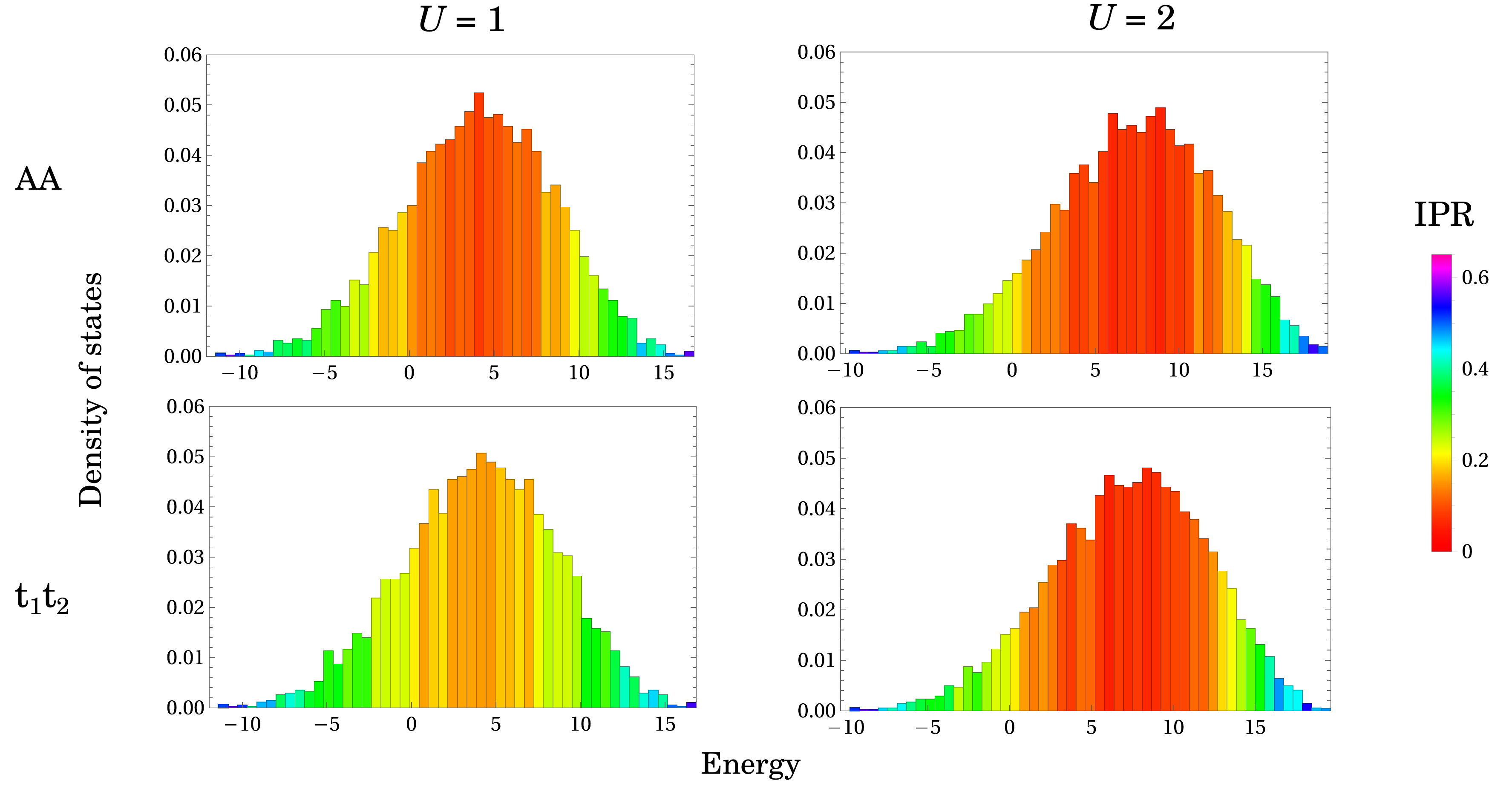}
	\caption{Density of states and energy-resolved IPR for the AA and the \modelname\ model.  
    As the interaction is turned on, the distinction between the AA and \modelname\ model spectra is suppressed and the two models become quite similar, with extended states concentrated in the mid-spectrum and localized states pushed to the edges.\label{figMBME}}
\end{figure*}

\section{Hilbert space path analysis and the many-body mobility edge \label{Appendix:F}}

\begin{figure*}[!]
	\includegraphics[width=0.8\textwidth]{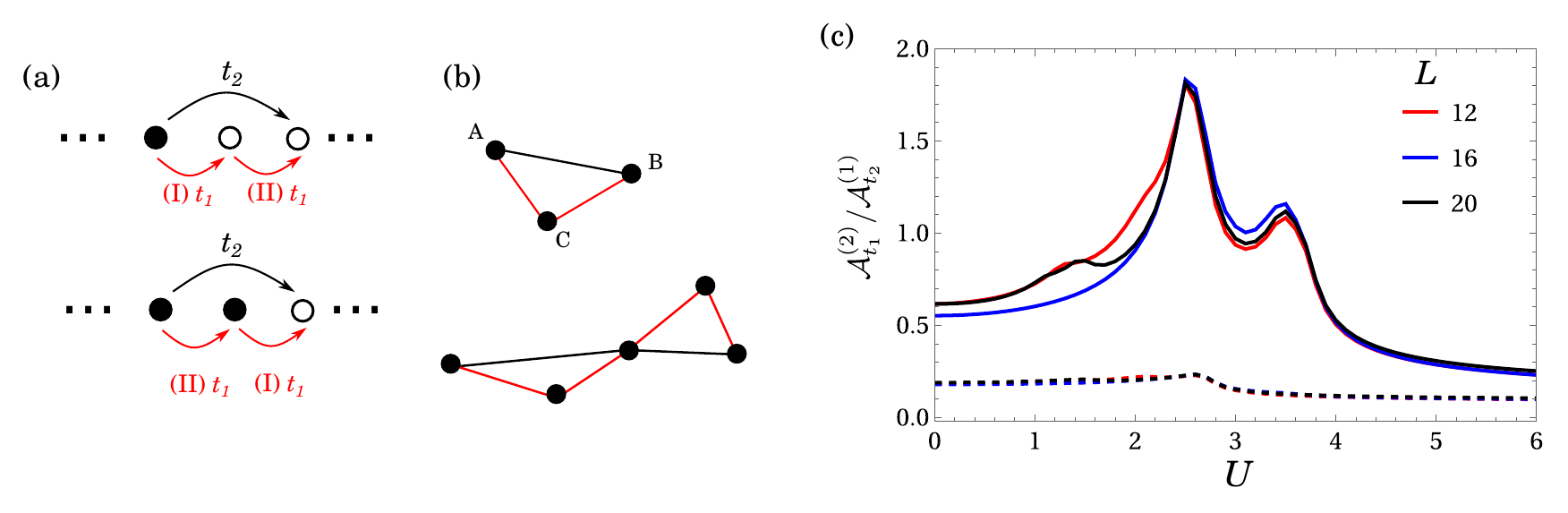}
	\caption{(a) The $t_2$-hopping process moves an electron by two sites and is equivalent to two consecutive $t_1$-hoppings. The empty/filled dots correspond to the respective empty/filled lattice sites. (b) In the Hilbert space, each link generated by $t_2$ can be substituted by two $t_1$ links. (c) Amplitude $\mathcal{A}_{t_2}^{(1)}$ (dashed lines) and $\mathcal{A}_{t_1}^{(2)}$ (solid lines) as a function of $U$ for $V=2$ and half-filled lattices. Different colors correspond to different system sizes. \label{path}}
\end{figure*}

\subsection{The many-body mobility edge in the AA and \modelname\ models} 
We now present the energy-resolved IPR spectrum for the \modelname\ model and the AA model with $L=14$, $N_e=7$, and $V=2.5$ in Fig.~\ref{figMBME}. 
As the interaction is turned on, there is a hybridization between noninteracting Slater determinant states, resulting in the spectrum being evenly covered and fitted to a Gaussian distribution. 
In addition, the energy-resolved IPR of the two models starts to converge into a universal form, i.e. extended states concentrated in the middle of the spectrum while localized states are pushed to the edges. 
This shows that the observed many-body mobility edge is universal and not tied to the single-particle mobility edge.

\subsection{Hopping strength under resonance}
In this subsection, we study the relative relevance of NNN hopping term $t_2$ compared to the nearest hopping $t_1$, which helps us better understand Fig. 2 in the main text. 
Given that the NNN hopping moves an electron by two lattice sites, such a process can also be substituted by two consecutive single-site hopping terms $t_1$ [see Fig.~\ref{path}(a)]. 
MBL can be studied in the Hilbert space map where each vertex is a Fock configuration and the link is generated by the hopping term \cite{Pietracaprina2016}. 
Accordingly, each link generated by $t_2$ has the same effect as two consecutive links generated by $t_1$ [see Fig.~\ref{path}(b)]. 
The amplitudes of links $A\to B$ and $A\to C \to B$ are respectively given by 
\begin{align}
	\begin{split}
		&\left|\mathcal{A}_{t_2}^{(1)}\right| = \frac{t_2}{\sqrt{(E_A-E_B)^2+\delta}}\\ 
		&\left|\mathcal{A}_{t_1}^{(2)}\right| = \frac{t_1}{\sqrt{(E_A-E_C)^2+\delta}}\frac{t_1}{\sqrt{(E_A-E_B)^2+\delta}}.
	\end{split}
\end{align} 
Numerically we fix $\delta=0.01$ to soften singularities caused by accidental degeneracy and average over all possible links and phases $\phi$ of the quasi-periodic potential $V_j = V \cos(2\pi qj+\phi)$. 
As shown in Fig.~\ref{path}(c), $|\mathcal{A}_{t_2}^{(1)}|$ only weakly depends on the interaction strength $U$ and decreases after $U>3$. 
On the contrary, $|\mathcal{A}_{t_1}^{(2)}|$ increases significantly when $U$ increases up to $3$ because it is more sensitive to interaction-induced resonances. 
Around this point, the NN hopping is much larger than the NNN hopping, rendering the latter irrelevant. 
For $U>4$, $\mathcal{A}_{t_2}^{(1)}$ and $\mathcal{A}_{t_1}^{(2)}$ stabilize and are comparable to each other, making the $t_2$-hopping process important again. 
This is qualitatively consistent with our results in the main text.

\bibliographystyle{apsrev4-2}
\bibliography{t1t2_MBL_v2}

\end{document}